\documentstyle[prl,aps,preprint,epsf]{revtex}
\begin{document}

\title{ Generalized stacking fault energy surfaces and 
dislocation properties of aluminum}

\author{Gang Lu, Nicholas Kioussis}
\address{Department of Physics,California State University Northridge,\\
 Northridge, CA 91330-8268}

\author{Vasily V. Bulatov}
\address{Department of Mechanical Engineering,\\ Massachusetts Institute of 
Technology, Cambridge, MA 02139}

\author{Efthimios Kaxiras}
\address{Department of Physics, Harvard University, Cambridge, MA 02138}
\maketitle

\begin{abstract}
We have employed the semidiscrete variational 
generalized Peierls-Nabarro model to study 
the dislocation core properties of aluminum. 
The generalized stacking fault energy 
surfaces entering the model are calculated by using 
first-principles Density Functional Theory (DFT) with pseudopotentials and 
the embedded atom method (EAM).  Various core properties, 
including the core width, 
splitting behavior, energetics and Peierls stress for different 
dislocations have been 
investigated. The correlation between the core energetics and 
dislocation character has 
been explored. Our results reveal a simple relationship between 
the Peierls stress and 
the ratio between the core width and atomic spacing. 
The dependence of the core 
properties on the two methods for calculating the total energy
(DFT vs. EAM) has been examined. 
The EAM 
can give gross trends for various dislocation properties but 
fails to predict the 
finer core structures, which in turn can affect the Peierls 
stress significantly (about 
one order of magnitude).
\end{abstract}

\begin{center}
{\bf \S 1. INTRODUCTION} 
\end{center}

Dislocations which are
one dimensional topological defects, are central to the understanding of mechanical properties 
of crystalline solids. The creation and motion of dislocations  
mediate the plastic response 
of a crystal to external stress. 
While continuum elasticity theory describes well the
long-range elastic strain of a dislocation for length scales beyond 
a few lattice spacings, it breaks down near 
the singularity in the region surrounding the dislocation center, 
known as the dislocation core. The discrete nature of the real 
crystalline lattice avoids the conceptual difficulty posed by the  
continuum singularity and recovers the structural differentiation 
smoothed out by the 
continuum elasticity.
There has been a great deal of interest in describing accurately  
the dislocation core structure on an
atomic scale because of its  important role in many phenomena 
of crystal plasticity
(Duesbery and Richardson 1991, Vitek 1992). 
The core properties control, for instance, the mobility of dislocations, which 
accounts for the intrinsic ductility or brittleness of solids.
The core is also important for the interaction of 
dislocations at close distances, which are relevant to plastic deformation. 
For example, by integrating the local rules derived 
from atomistic simulations of core interactions
into dislocation-dynamics simulations, a 
connection between micro-to-meso scales can be 
established to study dislocation reactions 
and crystal plasticity (Bulatov, Abraham, Kubin, Devincre and Yip 1998). 

Two types of theoretical approaches have been employed to 
study the core properties of dislocations. 
The first type is based on direct atomistic simulations 
employing either empirical potentials or 
first-principles calculations. 
Empirical interatomic potentials involve the 
fitting of parameters to a predetermined database
and hence may not be reliable in describing the 
core properties, where
severe distortions like 
bond breaking, 
bond formation and switching necessitate 
a quantum mechanical description of the electronic degrees of freedom.
On the other hand, first-principles electronic structure calculations, 
though considerably more accurate,  
are computationally 
expensive  for 
studies of  dislocation properties. 
The second type of approach is based on the framework of 
the Peierls-Nabarro (P-N) model which seems to be a plausible alternative
to direct atomistic simulations.  
In fact, there has been 
a resurgence of interest in the simple and tractable P-N model for the study of 
dislocation core structure and mobility 
(Schoeck 1994, Jo\'{o}s, Ren and Duesbery 
1994, Juan and Kaxiras, 1996).

Peierls first proposed the remarkable hybrid model (1940) in 
which some of the details of the discrete dislocation core 
are incorporated into an 
essentially continuum framework.  
Nabarro (1947) and Eshelby (1949) further developed 
Peierls' model and gave the first meaningful estimate 
of the lattice friction to 
dislocation motion. 
Later attempts to generalize the original treatment of Peierls and 
Nabarro assumed a more general core configuration from which they
derived 
the interactions between the glide planes which satisfy the Peierls 
integral equation. The essence of these models was captured in a more 
comprehensive approach by Vitek (1968, 1974), 
who introduced the concept of the generalized
stacking fault:
Consider a perfect crystal cut across a single 
plane into two parts which are then subjected 
to a relative displacement through an 
arbitrary vector $\vec{f}$ and rejoined. 
The reconnected lattice will have a surplus 
energy per unit area $\gamma(\vec{f})$. 
As the vector $\vec{f}$ is varied to span a 
unit cell of the interface, $\gamma(\vec{f})$ generates 
the generalized stacking fault (GSF) energy surface. 
The procedure can be repeated for 
various crystal planes. The significance of the GSF surface 
(or $\gamma$-surface) is 
that for a fault vector $\vec{f}$ there is an interfacial restoring stress 
\begin{equation}
\vec{F}_b(\vec{f}) = - \nabla (\gamma(\vec{f}))
\end{equation}
which has the same formal interpretation as 
the restoring stress in the P-N model.
The P-N model has now come to represent a combination 
of the original continuum model 
and the GSF interplanar potential, and its accuracy can be affected by either 
component. At present, the GSF energies can be calculated using empirical 
interatomic potentials (like the embedded atom method - EAM) 
or electronic structure methods.   
While 
extremely useful as a conceptual framework, the P-N model becomes 
increasingly inaccurate for dislocations with narrow cores, 
which is typically the case 
in covalently bonded solids 
(Jo\'{o}s, Ren and Duesbery 1994, Miller and Phillips 1996). 
The origin of this inaccuracy remains controversial, 
and it has not been unequivocally established whether or not 
the Peierls framework can be extended to capture such situations.  
Exploring the limits and extending the range of applicability of the 
classic P-N models remains a worthwhile endeavor. 
Recently, a 
semidiscrete  variational generalized P-N model has been proposed 
by Bulatov and 
Kaxiras (1997), which has been successfully implemented to a study of 
dislocation mobility in silicon. The new model predicts that the 
Peierls stress for the screw dislocation in Si is 0.065 eV/\AA$^3$ 
(using an energy surface from a classical interatomic potential), 
more than two orders of magnitude 
lower than the value 9.0  eV/\AA$^3$ obtained from 
the classic P-N model. 
A direct
atomistic calculation using the same interatomic potential gives
0.021  eV/\AA$^3$.
The new model also gives satisfactory results for
other core properties when compared to direct 
atomistic simulations for various dislocations in Si. 

The purpose of this paper is to apply this new model to
aluminum which is a prototypical ductile metal with much lower Peierls energy 
and stress than silicon. 
On the other hand, just like silicon, 
aluminum is 
known to have a narrow core due to its large stacking fault energy. 
The successful application of the new model 
to aluminum will further prove its validity 
and versatility in predicting dislocation core 
properties of different materials. 
We have  calculated the Peierls stress using both the 
semidiscrete generalized P-N model and 
atomistic simulations based on EAM. 
We have carried out systematic 
calculations of the core properties 
and the mobility of relevant dislocations in Al, and we have examined 
the relationship between the
core properties (energetics, core width and Peierls stress) and 
the dislocation 
character, namely the angle between the 
dislocation line and its Burgers vector. 
In order to explore the dependence of the dislocation properties on the 
method employed for the GSF energy calculations, we have 
performed calculations using 
both first-principles electronic structure 
methods and the empirical EAM potential.

The remainder of this paper is organized as follows: 
\S 2 describes the computational 
techniques used in our first-principles calculations for the GSF energy 
surface.  
\S 3 contains a brief review of the new model on which this study is based. 
In \S 4 we present the results of the GSF energy surfaces 
for the (111) plane of aluminum
using both the first-principles and empirical potential
calculations. In \S 5 we compare the dislocation 
properties using the GSF surfaces from the two methods for obtaining the 
energetics. 
The correlation between the dislocation properties 
and the dislocation character is presented in \S 6, along with 
some general conclusions on the applicability of our approach.

\begin{center}
{\bf \S 2. COMPUTATIONAL METHODS} 
\end{center}
The GSF energy surface is calculated within the framework of 
density functional theory (Hohenberg 
and Kohn 1964) in the local density density 
approximation (Kohn and Sham 1965) to the exchange-correlation functional, 
using the expression proposed by Perdew and 
Zunger (1984). We refer to these calculations by DFT in the following.
A kinetic energy cutoff of 12 Ry for the plane wave 
basis is used in the present calculations and
the atomic structures are considered fully relaxed when the Hellmann-Feynman
forces on each atom are smaller than 0.001 Ry/au.
The calculated equilibrium lattice constant and bulk modulus are 3.94 \AA~ 
and 82.52 GPa, in good agreement with the 
corresponding experimental room-temperature lattice 
constant and bulk modulus of 4.05 \AA~ 
and 76.93 GPa, 
respectively.  

In addition to the lattice constant and bulk modulus, we 
have calculated the value of the intrinsic stacking fault energy and 
the unstable stacking fault energy. 
To simulate the block shearing process we 
employed a supercell consisting 
of six or nine atomic layers in the (111) direction.
The intrinsic stacking fault configuration  corresponds to 
a slip of $a_0$/$\sqrt{6}$ in the $\langle$112$\rangle$ direction resulting in 
the stacking ABC$|$BCABC. The unstable stacking fault energy
corresponds to 
the lowest energy barrier that needs to be crossed for the slip 
from the ideal configuration to the intrinsic stacking 
fault in the $\langle$112$\rangle$ direction.
 The calculations are performed at the theoretically 
determined in-plane lattice constant to 
eliminate any artificially imposed stress.  
For the reciprocal space integration we have used a k-point
grid consisting of (16, 16, 4) divisions 
along the reciprocal lattice directions 
according to the Monkhorst-Pack scheme (Monkhorst and Pack 1976).
This corresponds to 514 k-points 
in the entire Brillouin zone.  Convergence tests were performed 
both for the number of divisions along each reciprocal space direction   
as well as the number of plane waves. Furthermore, the results for 
the 6-layer and 9-layer supercells are reasonably close, indicating 
adequate convergence with respect to the supercell size. 
Both atomic relaxations and volume relaxations 
were carried out to obtain accurate GSF energies. 
The value of the intrinsic stacking fault energy we obtained is 
 0.164 J/m$^2$, in excellent 
agreement with the result 
of  0.165 J/m$^2$ of Sun and Kaxiras (1997), 
and that of 0.161 J/m$^2$ of Wright, Daw and Fong (1992).  
 Experimental 
measurements range from a low value of 0.110 J/m$^2$ to a high value of 
0.280 J/m$^2$.  
Finally, the value 
of the relaxed unstable stacking fault energy
obtained from our calculations is 0.224 J/m$^2$, 
which is identical to the value obtained by Sun and Kaxiras (1997). 
 
\begin{center}
{\bf \S 3. SEMIDISCRETE VARIATIONAL PEIERLS-NABARRO MODEL} 
\end{center}
To facilitate the presentation we adopt the following conventions:
In Fig. 1, the XZ plane is
the (111) glide plane,  the Z axis is in the direction of the dislocation
line, and the X axis
is in the glide direction. The Y
axis in normal to the glide plane. For planar dislocations,  the displacements
along the Y direction are small. The Burgers vector $\vec{b}$ lies in the 
glide plane making an angle 
$\theta$ with the Z axis.
The Burgers vector is along the X axis ($\theta = 90^\circ$) 
for an edge dislocation 
and along the Z axis ($\theta = 0^\circ$) for a screw dislocation.
The Burgers vector of a mixed dislocation has both an edge 
component, $b \sin\theta$, 
and a screw component, $b \cos\theta$.
In general, the atomic displacements have components in
all three directions rather than only 
along the direction of the Burgers vector. This is because the path
along the Burgers vector may have to surmount a higher
interplanar energy barrier in the GSF surface, that is, 
the GSF energy is reduced when the
dislocation acquires additional displacement components in other directions.

In the classic P-N formalism, the dislocation misfit is assumed 
to be confined into a 
single plane, the glide plane, separating two 
semi-infinite linear elastic continua.
Between these two elastic half-spaces  is placed a dislocation,  conveniently
represented as a continuous distribution of 
infinitesimal dislocations (Eshelby 1949) 
with density $\rho (x)$. Here X is the coordinate of the atomic row, 
which is always parallel to the dislocation line.
A discrete lattice of arbitrary structure, deformed by
the dislocation's displacement field, is superimposed on 
the elastic half-crystals.
At a
given point along the interface the resultant misfit, 
due to all the infinitesimal
dislocations, is then balanced against the lattice restoring stress across the
glide plane, $F_b(f(x))$. 
This results  
 in the P-N integrodifferential equation
\begin{equation}
K\int\limits_{-\infty}^{+\infty} \frac{1}{x-x^\prime}
\frac{df(x^\prime)}
{dx^\prime} dx^\prime = F_b(f(x)).
\end{equation}
Here, $f(x)$ is the disregistry vector of the atomic 
row at point $x$ related to the dislocation density by 
$\rho(x) = df(x)/dx$,
and $K$ is a constant depending on the 
elastic properties and the dislocation character.
Its
value for an isotropic solid is given by:
\begin{equation}
K = \frac{\mu}{2\pi}(\frac{\sin^2\theta}{1-\nu} + \cos^2\theta),
\end{equation}
where $\mu$ and $\nu$ are the shear modulus and Poisson's ratio, respectively,
and $\theta$ is the angle between the dislocation line and the Burgers vector.
Thus, the pre-logarithmic elastic energy factor for a screw
dislocation is
 $K_s = \mu/2\pi$, while for an edge dislocation it becomes 
$K_e = \frac{\mu}{2\pi(1-\nu)}$.  
 The dislocation density
 $\rho (x)$ satisfies the normalization condition
\begin{equation}
\int\limits_{-\infty}^{+\infty}\rho(x^\prime) dx^\prime =
\int\limits_{-\infty}^{+\infty}\frac{df(x^\prime)}{dx^\prime} dx^\prime = b.
\end{equation}
If a simple sinusoidal form is assumed for F$_b(f(x))$, as in the original
P-N model, the disregistry vector is then given by the well-known analytical solution,
\begin{equation}
f(x) = \frac{b}{\pi}\tan^{-1}\frac{x}{\zeta} + \frac{b}{2}, 
\end{equation}
where 
\begin{equation}
\zeta = \frac{Kb}{2 F_{max}} 
\end{equation}
is the half width of the 
dislocation and F$_{max}$ is the maximum restoring stress. 

As pointed out recently by Bulatov and Kaxiras (1997), 
the classic P-N continuum model  
has the following flaws:
(1)While the elastic energy between the infinitesimal 
dislocations is evaluated from
a continuous integration, the misfit energy is sampled discretely
across the glide plane in order to incorporate the discrete nature of
lattice. Thus, the treatments of the two energy contributions are
not on equal footing and the total energy is not variational.
(2) The classic P-N model neglects the
important degrees of freedom which participate actively in the translation of a
dislocation over the Peierls barrier. 
(3) The 
elastic strain energy of a dislocation calculated within the P-N model can be
unrealistically high, especially for solids with a narrow core.  

In order to resolve these problems, the semidiscrete 
variational generalized Peierls model has been developed
by Bulatov and Kaxiras (1997).  
Within this approach the equilibrium structure of a dislocation is obtained 
by minimizing the dislocation energy functional
\begin{eqnarray}
U_{disl}&=& 
\sum_{i,j}\frac{1}{2}\chi_{ij}[K_e(\rho_i^{(1)}\rho_j^{(1)} + 
\rho_i^{(2)}\rho_j^{(2)}) + K_s\rho_i^{(3)}\rho_j^{(3)}] \nonumber \\  
& & 
+ \sum_i\Delta x \gamma_3(\vec{f_i})
- \sum_{i,l}\frac{x_i^2-x_{i-1}^2}{2}(\rho_i^{(l)}\tau_i^{(l)}) 
\label{ene_fun}
\end{eqnarray}
with respect to the dislocation density or disregistry vector.
Here, $\rho_i^{(1)}$, $\rho_i^{(2)}$ and $\rho_i^{(3)}$ 
are the edge, vertical and 
screw components of the general interplanar displacement density 
at the $i$-th nodal point, and $\gamma_3(\vec{f}_i)$ is the 
three-dimensional misfit potential.
The corresponding applied stress components interacting 
with the $\rho_i^{(1)}$, $\rho_i^{(2)}$ and 
$\rho_i^{(3)}$, are 
 $\tau^{(1)} = 
\sigma_{21}$, $\tau^{(2)} = \sigma_{22}$ and 
$\tau^{(3)} = \sigma_{23}$,
 respectively. $K_e$ and $K_s$ are the edge and screw 
pre-logarithmic energy factors defined earlier. 
The dislocation density at the $i$-th nodal point is 
$\rho_i = (f_i - f_{i-1})/(x_i - x_{i-1})$,
where $f_i$ and $x_i$ are the disregistry vector and 
the coordinate of the 
$i$-th nodal point (atomic row), respectively. 
The remaining quantities entering in this expression are: 
$\chi_{ij} = \frac{3}{2}\phi_{i,i-1}\phi_{j,j-1} + 
\psi_{i-1,j-1} + \psi_{i,j} 
- \psi_{i,j-1} - \psi_{j,i-1}$, with $\phi_{i,j} =  
x_i - x_j$, and $\psi_{i,j} = 
\frac{1}{2}\phi_{i,j}^2\ln\vert\phi_{i,j}\vert$.

The first term in the energy functional, 
Eq.(7), represents the elastic energy, which
has been discretized. Since 
any details of the displacements across the slip plane other 
than those on the atomic rows 
are disregarded, the dislocation density is constant between the 
nodal points.
This  explicit discretization of the elastic energy 
term removes the inconsistency
in the original P-N model and allows the total energy functional to be
variational.
Another modification in this approach is 
that  the nonlinear misfit potential in the energy functional is a function 
of all three components of the nodal displacements, $\vec {f}(x_i)$.
Namely,  in addition to the displacements along the Burgers vector, 
lateral and even vertical displacements across 
the slip plane  
are also included.
This in turn allows the treatment of straight 
dislocations of arbitrary orientation in arbitrary glide planes. 
Furthermore, because the disregistry vector $\vec{f}(x_i)$ is allowed 
to change during the process of dislocation translation, 
the Peierls energy barrier  
can be significantly lowered compared to the corresponding 
value from a rigid translation.    
The response of the dislocation to an applied stress is 
achieved by minimization of
the energy functional with respect to $\rho_i$ at 
a given value of the applied stress, $ \tau_i^{(l)}$.  
An instability is reached 
when an optimal solution for $\rho_i$ no longer exists, which 
is manifested numerically by the failure of the minimization 
procedure to convergence. 
Within this formulation, we take as the definition of the Peierls stress
the critical value of the external applied stress  
which produces this instability. 

\begin{center}
{\bf \S 4. GENERALIZED STACKING FAULT ENERGIES} 
\end{center} 

The first-principles 
GSF energy surface $\gamma(\vec{f})$ 
for the (111) plane was calculated 
on a dense grid of 40 points in the 
irreducible part of the (111) slip plane 
(which is 1/12 of the area shown in Fig. 2). We 
have used a symmetrized polynomial basis to fit the 
GSF energy surface in order to 
evaluate the restoring force and facilitate the computation of dislocation 
properties. Since it is relatively much faster to calculate the GSF
energy using the EAM method, we directly compute
the GSF energy for any given disregistry vector.
Although we do not need the EAM GSF energy surface explicitly
for calculating any dislocation properties, we also fit that energy surface
with the same basis of symmetrized polynomials for the purpose of comparison 
to the DFT calculation.  The two energy surfaces are shown in Fig. 2. 
 
Both GSF surfaces maintain the rotation 
symmetry of the fcc lattice.
The three high peaks of the GSF 
surfaces correspond
to the run-on stacking fault configuration  
ABC$|$CABC, in which the two C layers are  
neighboring to each other.  
The projection of the GSF energy surfaces onto the [12$\bar{1}$] and [101] 
directions are 
shown in Fig. 3(a) and 3(b), respectively. 
Along the [12$\bar{1}$] direction, 
the first energy maximum
encountered is the 
unstable stacking fault energy, which 
represents the lowest energy barrier for dislocation nucleation
(Rice 1992, Rice and Beltz 1994)
and the first energy  
minimum at $a_0$/$\sqrt{6}$ (where $a_0$ is the lattice constant)
corresponds to the intrinsic stacking fault configuration, 
where 
a full dislocation dissociates into a pair of Shockley partials. 
On the other hand, the GSF surface projection along the [101] direction 
is symmetric with respect to the slip $a_0$/$2\sqrt{2}$.
For both the DFT and EAM calculations, 
the unstable stacking fault energy along [101] 
is found to be larger than that along [12$\bar{1}$].
This anisotropy of the unstable stacking fault energy will affect 
the emission of
dislocations from a crack tip.
The energy values for the various stacking faults 
obtained from the DFT and EAM calculations are summarized
in Table I.
The restoring stresses from both calculations and for different directions are 
shown in Fig. 4(a) and 4(b), respectively.  

\begin{center}
{\bf \S 5. DISLOCATION PROPERTIES}  
\end{center}
 
In this section we present the results of various 
dislocation properties for the screw, 
30$^\circ$, 60$^\circ$ and edge dislocations with 
the Burgers vector, $\vec{b}$ = a/2 [101],  using the GSF 
energy surface obtained from the DFT calculations and the 
direct EAM energy 
calculations.   

\begin{center}
{\bf \S 5.1. Disregistry vector and dislocation width} 
\end{center}  

The disregistry vector $\vec{f}=f_1\hat{x}+f_2\hat{y}+f_3\hat{z}$ for
the four dislocations can be calculated from the model using both the 
DFT and EAM energy surfaces (see Fig. 1 for the definition
of X,Y and Z). Since all four dislocations have the same Burgers vector
$\vec{b}$ but different orientations (different angle of
the dislocation line with
respect to $\vec{b}$), in order to examine the 
trend as a function of the angle one needs to 
project the disregistry vector $\vec{f}$ in two directions, parallel 
and perpendicular to the Burgers vector $\vec{b}$. 
The components of $\vec{f}$ parallel and perpendicular to  
the Burgers vector are $f^a=f_1\sin\theta + f_3\cos\theta$ and 
$f^b=-f_1\cos\theta+f_3\sin\theta$,
respectively.
The results for $f^a$ from both energy surfaces for the screw
and 60$^\circ$ dislocations are presented in Fig. 5(a) 
and for the 30$^\circ$ and edge dislocations are
shown in Fig 5(b) (all in units of $|\vec{b}|$).

In order to examine the degree of deviation of the disregistry vector 
$\vec{f} (f^a, f^b)$
from the direction of the Burgers vector $\vec{b}$, we present in Fig. 6
the disregistry path for the four dislocations 
using the DFT (Fig. 6(a))
and EAM results (Fig. 6(b)).
Although both methods predict that for the 
four dislocations the disregistry paths $\vec{f}_i (f^a_i, f^b_i)$ 
deviate significantly from the direction of 
$\vec{b}$ (vector $\vec{AB}$ in the figures) in order to lower the total
energy, the magnitude of deviation obtained from 
the DFT calculation is
much smaller than that from the EAM energy surface. 
More importantly, the deviation from
the Burgers vector is almost zero in Fig. 6(a) for
the atomic rows at the center of
the dislocations  
whereas it reaches a maximum in Fig. 6(b). Apparently the difference between
the two GSF energy surfaces results in the different displacement paths for the
four dislocations. 

One can determine from Fig. 5 the dislocation half width $\zeta$,  defined
as the atomic distance (nodal spacing) over which $f^a$ changes
from (1/4)$|\vec{b}|$ to (3/4)$|\vec{b}|$. 
The half width can also be calculated from 
the classical P-N model which assumes a sinusoidal form for the
restoring stress $F_b(x)$ (Eq. (6)).  Comparison of these two 
definitions provides information on how important 
are the details of the restoring stress in determining $\zeta$.
We have used a sinusoidal restoring stress with F$_{max}$ = $F_{us}$, 
where $F_{us}$ is defined as the first maximum restoring stress encountered
along the [12$\bar{1}$] direction for both the DFT and EAM cases. 
Values of the dislocation width of the four dislocations 
evaluated from the above two definitions
are listed in Table II.
We find that both DFT and EAM calculations give 
a dislocation core width that increases monotonically with
the dislocation angle.  The EAM calculation gives a wider dislocation 
core compared to that obtained from the DFT energy surface.  
This is due to the fact that 
the EAM potential gives smaller restoring stresses and GSF energies for 
a wide range of
shears along the [12$\bar{1}$] direction, and in particular for the
$F_{us}$ and the intrinsic 
stacking fault energy $\gamma_{is}$ 
(see Fig. 3(a) and 4(a)).  
The smaller $F_{us}$ gives rise to a
larger core width, while the smaller $\gamma_{is}$  
results in the dissociation of the full dislocation into
partials and hence a wider core.

It is important to note that overall there
is a good agreement (in particular for the DFT 
calculations) for the core width 
as evaluated from the two
definitions described above.  
While the first definition takes into account
the full details of the entire GSF surface in the evaluation of the 
disregistry vectors, the second definition 
involves only the maximum restoring stress $F_{us}$.
The agreement suggests that the details of the GSF surface 
are not important as far as the core width
 is concerned.
The difference between the two definitions
is larger  
for the 60$^\circ$ and edge dislocations
when the EAM potential is used.  This is because 
the EAM energy surface gives rise to a splitting into partials 
for both dislocations, which could only be manifested within the first
definition.

\begin{center}
{\bf \S 5.2. Splitting into the partials} 
\end{center}

The dislocation density for the screw, 
30$^\circ$, 60$^\circ$, and edge 
dislocations calculated from the DFT GSF energy surface
(solid curves) are
presented in Fig. 7, 
together with the corresponding results from the EAM calculation
(dashed and dotted curves).
One can see that 
in all cases 
the DFT energy surface gives no splitting 
of the complete dislocations into partials. 
While the narrow double-peak structure found for the 60$^\circ$ complete 
dislocation is suggestive of a splitting, it is rather due  to the fact that 
the nodal points along the X direction are not evenly spaced, 
i.e. they are distributed 
alternately by $\vec{b}_p$ and $\vec{b}_p$/2, where $\vec{b}_p$ is the 
Burgers vector of the Shockley partial.
This in turn gives rise to density fluctuations over the neighboring 
atomic rows and hence a double-peak structure.
A double-peak structure may indicate 
a splitting if the peaks are separated by a larger distance or the 
nodal points are evenly spaced, as in the cases of the
 30$^\circ$ and edge dislocations.  

As far as these results are concerned, 
for the 30$^\circ$ and the 60$^\circ$ dislocations we plot both
the edge and screw components.
In contrast to the DFT results,   
the EAM calculations predict that 
the full edge and  60$^\circ$ 
dislocations will split into partials.
The multi-peak found for the screw-dislocation suggests
it might also be unstable. The 30$^\circ$  
dislocation exhibits a {\it narrow} double-peak structure.
The overall splitting trend found in the EAM calculations is due to
the fact that the EAM potential gives a smaller intrinsic stacking
fault energy. Realizing that the EAM GSF energy surface differs from the
DFT surface not only in the intrinsic stacking fault (ISF) energy, but
also the energy profile around it, we have investigated the effect of
the shape of the GSF surface on the splitting behavior for the full edge
dislocation using the DFT GSF surface. We have kept the ISF energy of
the DFT surface to be the real value of 0.164 J/$m^2$, but varied the
energy profile around the ISF to make it as flat as possible to
be alike the EAM GSF surface (Fig. 8). We then calculated the
dislocation density for the edge dislocation using the modified GSF
surface, and we find that there is a small dissociation into partials
(double-peak in the density distribution) with the spacing of one
Burgers vector between the partials (Fig.
9). Therefore not only the ISF energy itself, but also the energy
profile of the GSF energy surface around it could determine a splitting
into partials.
It is interesting to note that the EAM calculations predict a
separation between partials of about 8-9 \AA~
(about three 
times the Burgers vector), in agreement with the values 
of about 6-7 \AA~ predicted by the elastic continuum theory
 (Hirth and Lothe 1982).
It is also interesting to examine the character of the 
resultant partials.
The complete edge dislocation dissociates into two symmetric 60$^\circ$ 
partials, whereas the 60$^\circ$ dislocation dissociates into a 30$^\circ$ 
(left double-peak) and a 90$^\circ$ (right double-peak) partial.
The double-peak structure again stems from the inequivalent nodal 
spacing between neighboring atomic planes. The 
fact that the density of the screw 
component (dotted curve) vanishes at the point where 
the edge component (right double peak) reaches 
its maximum, indicates the pure edge character of the 90$^\circ$ partial.
The 30$^\circ$ dislocation 
exhibits a weak tendency for dissociation into a screw and
a 60$^\circ$ partial, while 
the edge component for the 60$^\circ$ partial is singly peaked.
Our results for the dissociation of the full 
60$^\circ$ dislocation into partials are in agreement 
with those from the direct atomistic simulations by Bulatov (1998)  
using the same EAM potential.  

\begin{center}
{\bf \S 5.3. Energetics and lattice resistance} 
\end{center}

In Table II we present the results for the
the core energy, defined as the sum of the elastic and misfit 
energies, the 
separate contributions to the core energy from the elastic  
and GSF energies,  
and the Peierls stress for the 
four dislocations using the DFT and EAM potentials.
Also for comparison, we include in Table II the values of
the Peierls 
stress for the screw and 60$^\circ$ dislocations 
obtained from the direct atomistic simulations 
using the same EAM potential. 

The results for the energetics from the DFT calculations 
are overall in good agreement with those from the EAM calculations, 
including the trend across 
the types of dislocations. 
We identify the configuration dependent part of the total
energy as the core energy, including the elastic energy and GSF energy.
The GSF energy contribution (positive in sign) increases monotonically in going 
from the screw to the edge dislocation; the elastic contribution 
(negative in sign) decreases as the angle of the Burgers vector with
respect to the dislocation line increases.  The elastic energy, 
ignored in some previous studies, is the  
dominant contribution to the core energy (about a factor of two larger than 
the GSF energy), and more importantly 
it is strongly dependent on the dislocation character.

The mobility of dislocations is characterized 
by the Peierls stress and the Peierls 
energy, the former being defined usually as 
the maximum derivative of the latter 
with respect to dislocation translation.
As mentioned earlier, 
the Peierls stress in this work is obtained by evaluating the 
critical value of the applied stress $\tau$, at which the 
dislocation energy functional fails
to be minimized with respect to $\rho_i$ through 
standard conjugate gradient techniques.  This approach is more 
accurate and physically transparent because it captures the nature of
the Peierls stress as the stress at which the displacement field of the 
dislocation undergoes a 
discontinuous transition.   

The results for the Peierls stress, listed in Table II,  
calculated from the DFT GSF energy surface are in good agreement with 
those from the direct EAM atomistic simulations and those from the 
EAM calculations except for the edge dislocation. 
For the edge dislocation, 
the EAM calculations yield a Peierls stress which is an order of magnitude
larger than that from the DFT calculations. 
This discrepancy could be due to the dissociation of the full edge 
dislocation into partials. The effects of the
dissociation on the Peierls stress have been studied
by Benoit {\it et al.} (1987).  
These authors argued that if the equilibrium separation 
between the two Shockley partials is an integer or half integer multiple of 
the vector 
$(a/2)\langle110\rangle$, so that the two partials move in phase and 
reach the troughs or the crests of the Peierls potential 
simultaneously, then the 
Peierls stress required to move the extended configuration 
is simply the stress 
required to move an isolated partial dislocation. 
On the other hand, if the two partials are 
rigidly separated by (1/4) or (3/4) times 
the vector $(a/2)\langle110\rangle$, then the partials are 
exactly out of phase and the Peierls stress 
on them are always equal and opposite, 
and hence the applied stress required 
to move the rigid configuration vanishes.

In order to understand the discrepancy 
for the Peierls stress of the edge dislocation from 
the two calculations, we next investigate the effects of the 
splitting on the Peierls stress. Since the EAM gives 
a dissociation for the perfect edge 
dislocation into two symmetric 60$^\circ$ partials with 
a separation of about three times the Burgers vector (see \S 5.2),
one can infer from the work of Benoit {\it et al.} that the Peierls 
stress for each of the two partials is equal to 
the Peierls stress required to move the extended dislocation, 
i.e. the two partials and the intrinsic 
stacking fault in between. 
Continuum elastic theory gives that the separation of the partials is
\begin{equation}
d = \frac{\mu b_p^2}{8\pi\gamma_{is}}\frac{2+\nu}{1-\nu},
\end{equation}
where $b_p$ is the partial Burgers vector and $\gamma_{is}$ is the intrinsic
stacking fault energy.
Since the DFT GSF energy surface predicts no splitting for the edge
dislocation, one can force the dissociation into partials by either increasing
the elastic constant $K$ in Eq. (3) 
or reducing the intrinsic stacking fault energy.

We find that increasing the pre-logarithmic factor $K$ 
in the elastic 
energy term (by up to a factor of 10) but keeping the same DFT GSF 
surface gives rise to a wider dislocation density distribution but 
yields no dissociation.
On the other hand, if we keep the same pre-logarithmic factor $K$ 
but reduce the DFT value for the intrinsic stacking fault $\gamma_{is}$
from 0.164J/m$^2$ to 0.096
J/m$^2$ (the vicinity of 
the stacking fault in the GSF surface is also reduced 
by this rescaling of the energy), 
we find that the perfect edge dislocation splits 
into two 60$^\circ$ partials separated by one Burgers vector and 
a Peierls stress of 0.33 meV/\AA$^3$, an order of 
magnitude larger than the original value of 0.02 meV/\AA$^3$. 
If we further reduce the stacking fault energy to 
0.085 J/m$^2$,  we find that the edge 
dislocation splits into two partials separated by a distance of 3$b$, 
but with the same Peierls 
stress (0.33 meV/\AA$^3$)
for the extended configuration.
Since our calculations give separations between partials that 
are integer multiples of the Burgers vector, we can infer from 
the work of Benoit {\it et al.} 
that the stress of 0.33 meV/\AA$^3$ should be equal to the 
Peierls stress for moving an isolated 60$^\circ$ partial. 
Thus, we find that the origin of the   
discrepancy of the Peierls stress for the edge dislocation between 
the DFT and EAM calculations  
is that the DFT calculations predict no 
splitting for the edge dislocation and hence a very 
low Peierls stress,  
whereas the EAM calculations predict a splitting  
 into partials due to the smaller stacking fault energy, and hence  
a much higher Peierls stress. 
It was not possible to test whether the Peierls stress 
for an isolated 60$^\circ$ partial is equal to
the Peierls stress for the extended dislocation (0.33 meV/\AA$^3$),  
because an isolated partial dislocation introduces an 
infinite stacking fault on the glide plane.
  
In order to 
test the accuracy of the present model for 
predicting the Peierls stress, we also have 
performed direct atomistic simulations using the same EAM
potential 
 for both the screw and 
60$^\circ$ dislocations.
The values for the Peierls stress from the direct atomistic 
calculations listed in Table II are in very good 
agreement with our model calculations. 
For comparison, we have also calculated 
the Peierls stress for the screw and 60$^\circ$ 
dislocations using   
the expression of Jo\'{o}s and Duesbery 
(1997) 
\begin{equation}
\sigma_p=\frac{Kb}{a'}\exp(-\frac{2\pi\zeta}{a'}),
\end{equation}
where  
$a'$ is the atomic spacing, while 
the core width $\zeta$ is determined from 
the EAM 
calculations. 
Eq. (9) gives a Peierls stress   
1.44$\times 10^{-2}$ meV/\AA$^3$ 
for the screw dislocation,
and of 4.8$\times 10^{-4}$ meV/\AA$^3$ for the 60$^\circ$ dislocation,
both several orders of magnitude smaller than the corresponding 
values found from the current model and the results of the direct 
atomistic simulations.

\begin{center}
{\bf \S 6. CORRELATION BETWEEN DISLOCATION PROPERTIES AND DISLOCATION CHARACTER}
\end{center}
Using the DFT GSF surface we have studied the 
dislocation properties of nineteen different dislocations 
that have the same Burgers vector 
but different orientations, 
 in order to examine the possible  
 correlation between the dislocation properties and the dislocation 
character. 
The angle between the 
dislocation line and the Burgers vector 
has been varied from 0$^\circ$ to 90$^\circ$. 

In Fig. 10 we present the dislocation core width as a function of
the dislocation angle using the two definitions described in \S 5.1.
The width increases monotonically with the dislocation angle.
Even though the first definition takes into account the details of the
entire GSF surface while the second definition employs only a single 
special point of a simpler sinusoidal GSF (namely $F_{us}$), 
the overall agreement is very good (less than 10\% difference).

The core energy, along with its separate energy contributions 
from the elastic energy 
(interaction 
between the infinitesimal parallel dislocations on the glide plane) 
and the GSF energy are presented in Fig. 11  
as a function of the angle. 
We find that the core energy and the elastic energy
decrease monotonically as the angle increases from 
0$^\circ$ (screw) to 90$^\circ$ (edge), whereas 
the GSF energy increases with the angle. 
As the angle increases 
the pre-logarithmic factor $K$ of a mixed dislocation increases and hence 
the elastic energy is lowered. On the other hand, the monotonic increase of
the GSF energy with 
respect to the angle is due to fact that the core width 
increases monotonically 
with respect to the dislocation character. 
The elastic energy dominates 
(in absolute value) the GSF energy and 
increases in magnitude faster with increasing angle than 
the GSF energy, 
indicating a weaker anisotropy for the 
GSF energy compared to that for the elastic energy. 
Thus, the elastic energy not only 
is the dominant contribution to the total energy stored in 
the core region, but also is
more sensitive to the dislocation character  
than the GSF energy.  

Finally, we have calculated the Peierls stresses for all the nineteen
dislocations
in an effort to correlate the Peierls stress with the dislocation 
character. 
Most of the dislocations in the fcc lattice have non-even 
nodal spacings, 
 except for the 
 30$^\circ$ and edge dislocations. 
In Fig. 12 we plot ln$(\sigma_p\bar{a}/Kb)$
as a function of $\zeta/\bar{a}$, where $\zeta$ is the half
width of the dislocation core calculated from the atomic
spacing over which the disregistry vector changes from
$|\vec{b}|/4$ to $(3/4)|\vec{b}|$, 
and $\bar{a}$ is the average nodal spacing along the X direction.
We also show for comparison 
the analytic expression Eq. (9) of
Jo\'{o}s and Duesbery (1997), 
with a dashed line (slope of -2$\pi$).       
It should be emphasized that this expression 
is valid only for evenly spaced atomic planes.
Most of
 the calculated values can be fitted with  
\begin{equation}
\sigma_p = \frac{Kb}{\bar{a}}e^{-1.7\zeta/\bar{a}}, 
\end{equation}
shown as the solid line.
The points corresponding to the 30$^\circ$ and edge dislocations 
deviate from the above line because of the even nodal spacing. 
We have recently formulated 
a general theory to account for the non-evenly spaced
dislocations (Lu, Kioussis, Bulatov and Kaxiras, 1999) which 
shows that the Peierls stress for evenly-spaced dislocations
is lowered by several orders of magnitude compared to that 
of the non-evenly spaced dislocations. 
On the other hand, the deviation from the common trend 
of the values for the 
10.9$^\circ$ and 14.9$^\circ$ dislocations is unclear at present. 
The Peierls stress is more sensitive to 
the average atomic spacing $\bar{a}$ 
than to the half width.
For example, while both the screw 
and 14.9$^\circ$ dislocations have predominant 
screw components and similar 
half widths of 2.1 \AA~ and 
2.3 \AA~, respectively, 
they have quite different atomic spacings, 
1.2 \AA~ and 
0.3 \AA~, respectively.  
This results in a Peierls stress of 0.04 meV/\AA$^3$ for the 
14.9$^\circ$ dislocation,
almost two orders of magnitude smaller than 
that of 1.60 meV/\AA$^3$ for the screw 
dislocation. 

In conclusion, 
we have performed DFT and EAM calculations to obtain the GSF 
energy surfaces for the (111) glide plane of Al.   
From those calculations we extracted the properties 
for various dislocations, using 
the semidiscrete variational generalization of the Peierls-Nabarro model.  
We have demonstrated that although the EAM gives the general trends
for various dislocation properties, 
it fails to predict the finer structure of the dislocation core, 
such as the dissociation into partials, which in 
turn determines the mobility of dislocations.
Since the splitting of a full dislocation into partials 
depends strongly on the intrinsic stacking fault energy, 
direct atomistic simulations based on empirical potentials
may also fail to predict the correct dissociation behavior. 
Accordingly, the results of the present work indicate that 
the accurate DFT calculations of the GSF surface 
combined with the semidiscrete variational P-N model enable 
the study of dislocation core properties 
more accurately and expediently. Moreover,  
this model could be extended to study a wide
range of problems that are associated 
with more complex dislocations processes such as cross slip, dislocation
intersection, reaction, etc.

\begin{center}
{\bf ACKNOWLEDGEMENTS} 
\end{center}
The work at California State University Northridge 
was supported by the grant No. 
DAAG55-97-1-0093 through the U.S. Army Research Office.
The work of EK is supported by the Harvard Materials Research 
Science and Engineering Center, 
which is funded by NSF grant number DMR-94-00396. 

\newpage

\begin{center}
REFERENCES
\end{center}
BENOIT, W., BUJARD, M., AND GREMAUD, G., 1987, 
{\it Phys. Status Solidi.}, {\bf A104}, 427. \\  
BULATOV, V., 1998, Private communication.\\
BULATOV, V., ABRAHAM, F. F., KUBIN, L., DEVINCRE, B., 
AND YIP, S., 1998, {\it Nature}. 
{\bf 391}, 669.\\
BULATOV, V. V., AND KAXIRAS, E., 1997, {\it Phys. Rev. Lett}., {\bf 78}, 4221.\\
DUESBERY, M. S., AND RICHARDSON, G. Y., 1991, 
{\it Solid State and Materials Sciences}, 
{\bf 17}, 1.\\
ESHELBY, J. D., 1949, {\it Phil. Mag.}, {\bf 40}, 903.\\
HIRTH, J. P., AND LOTHE, J., 1992, 
{\it Theory of Dislocations}, 2nd ed., (Wiley, New York).\\
HOHENBERG, P., AND KOHN, W., 1964, {\it Phys, Rev.}, {\bf 136}, B864.\\
JOOS, B., AND DUESBERY, M. S., 1997, {\it Phys. Rev. Lett}., {\bf 78}, 266.\\
JOOS, B., REN, Q., AND DUESBERY, M. S., 1994, 
{\it Phys. Rev}. B, {\bf 50}, 5890.\\
JUAN, Y., AND KAXIRAS, E., 1996, {\it Phil. Mag.} A, {\bf 74}, 1367.\\
KOHN, W., AND SHAM, L., 1965, {\it Phys. Rev.}, {\bf 140}, A1133.\\
LU, G., KIOUSSIS, N., BULATOV, V. AND KAXIRAS, E., 1999, to be published.\\
MILLER, R., AND PHILLIPS, R., 1996, {\it Phil. Mag}. A, {\bf 73}, 803.\\
MONKHORST, H. J., AND PACK, J. D., 1976, {\it Phys. Rev.} B, {\bf 13}, 5188.\\
NABARRO, F. R. N., 1947, {\it Proc. Phys. Soc. London}, {\bf 59}, 256.\\
PEIERLS, R., 1940, {\it Proc. Phys. Soc. London}, {\bf 52}, 34.\\
PERDEW, J., AND ZUNGER, A., 1984, {\it Phys. Rev.} B., {\bf 23}, 5048.\\
RICE. J. R., 1992, {\it J. Mech. Phys. Solids}, {\bf 40}, 239.\\
RICE, J. R., AND BELTZ, G. E., 1994, 
{\it J. Mech. Phys. Solids}, {\bf 42}, 333.\\
SCHOECK, G., 1994, {\it Phil. Mag}. A, {\bf 69}, 1085.\\
SUN, Y., AND KAXIRAS, E, 1997, {\it Phil. Mag.} A, {\bf 75}, 1117.\\
WRIGHT, A., DAW, M. S., AND FONG, C. Y., 1992, 
{\it Phil. Mag.} A, {\bf 66}, 387.\\
VITEK, V., 1968, {\it Phil. Mag}., {\bf 18}, 773.\\
VITEK, V., 1974, {\it Crystal Lattice Defects}, {\bf 5}, 1.\\
VITEK, V., 1992, {\it Progress in Materials Science}, {\bf 36}, 1.\\

\newpage

\begin{table}
\caption{The energetics and fault vectors for the 
four stacking faults obtained from the DFT  
and EAM calculations.  All energies in J/m$^2$.}
\begin{tabular}{cccc}
 &{Vector} & DFT  &  EAM \\ \hline
 Intrinsic stacking & 1/6[12$\bar{1}$]  & 0.164 & 0.120 \\ 
 Unstable stacking  & 1/10[12$\bar{1}$] & 0.224 & 0.141 \\
 Unstable stacking  & 1/4[101]          & 0.250 & 0.663 \\
 Run-on stacking    & 1/3[12$\bar{1}$]  & 0.400 & 1.354\\ 
\end{tabular}
\end{table}

\begin{table} 
\caption{Core widths $\zeta$ (in \AA~), with 
$\zeta^I$ from the 
definition through the disregistry vector and $\zeta^{II}$ from 
the definition through the restoring stress; 
total core energies and separate contributions 
from the elastic 
and GSF energies (in eV/\AA); 
and Peierls stress (in meV/\AA$^3$), 
for the four dislocations from both the DFT 
and EAM calculations. 
The last two rows list the Peierls stress 
for the screw and 60$^\circ$ dislocations 
from the direct atomistic simulations (At. Sim.) using the same EAM potential
and from the expression of Jo\'{o}s and Duesbery (JD).}
\begin{tabular}{cccccc}
               &     & screw     & 30$^\circ$& 60$^\circ$& edge      \\ \hline
Core widths $(\zeta^{I},\zeta^{II})$    
               & DFT  & (2.1,2.1)& (2.5,2.4) & (3.0,2.9) & (3.5,3.2) \\ 
               & EAM  & (3.7,3.4)& (4.3,3.8) & (5.4,4.7) & (6.4,5.2) \\ \hline 
Core energy    & DFT & $-0.0834$ & $-0.1096$ & $-0.1678$ & $-0.1979$ \\ 
               & EAM & $-0.0534$ & $-0.0813$ & $-0.1413$ & $-0.1730$ \\ \hline
Elastic energy & DFT & $-0.1828$ & $-0.2317$ & $-0.3199$ & $-0.3666$ \\
               & EAM & $-0.1679$ & $-0.2086$ & $-0.2960$ & $-0.3445$ \\ \hline
GSF energy     & DFT &   0.0938  &   0.1221  &  0.1521   &   0.1688 \\
               & EAM &   0.1145  &   0.1273  &  0.1547   &   0.1716 \\ \hline
Peierls stress & DFT & 1.60   & 0.33   & 0.61   & 0.02 \\
               & EAM & 0.55   & 0.21   & 0.28   & 0.15 \\
               & At. Sim. & 0.51   &   &0.29    &         \\
               & JD  & 0.0144  &   &0.00048 &  \\
\end{tabular}
\end{table}

\newpage

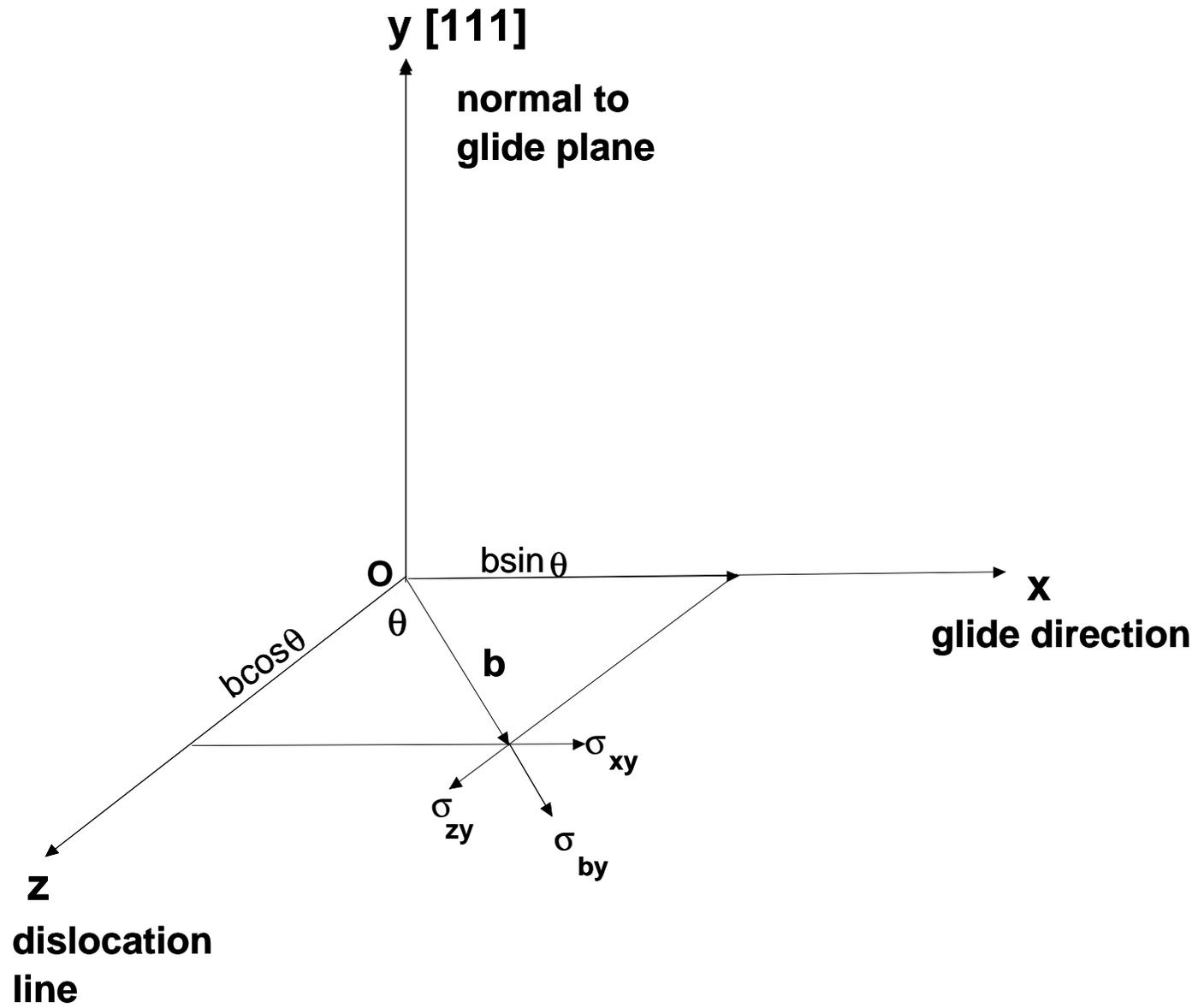
\begin{figure}
\caption{A Cartesian set of coordinates showing 
the directions relevant for dislocations 
in Al.} 
\label{fig1}
\end{figure}

\begin{figure}
\caption{The GSF energy surfaces for displacements along a (111) plane in Al
(J/$m^2$) (the 
corners of the plane and its center correspond to 
identical equilibrium configurations, 
i.e., the ideal Al lattice): 
(a) from DFT pseudopotential plane-wave 
calculations; (b) from EAM calculations.  The two energy surfaces are displayed
in exactly the same perspective and on the 
same energy scale to facilitate comparison of important features.
The EAM energy surface extends to values about 4 times larger than the 
maximum value displayed (black truncated regions).}
\label{fig2}
\end{figure}

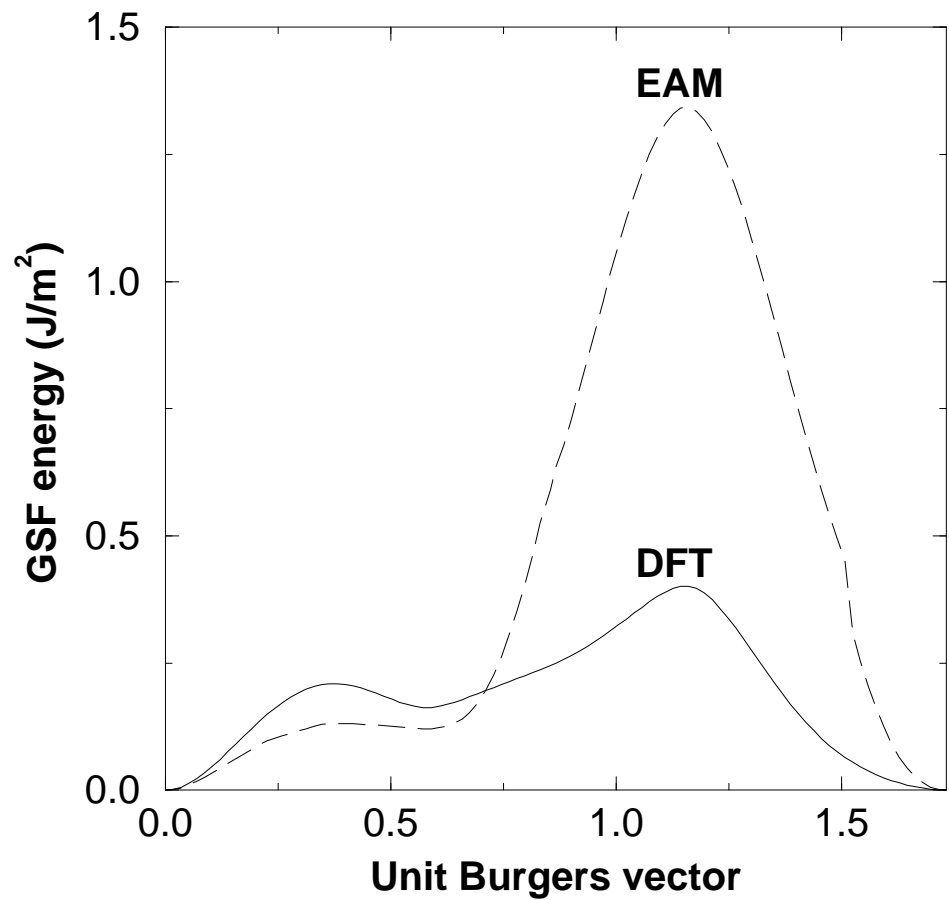
\begin{figure}
\caption{The projections of the GSF energy surfaces 
onto (a) [12$\bar{1}$] and (b) [101] directions from both the DFT and
EAM calculations.}
\label{fig3}
\end{figure}

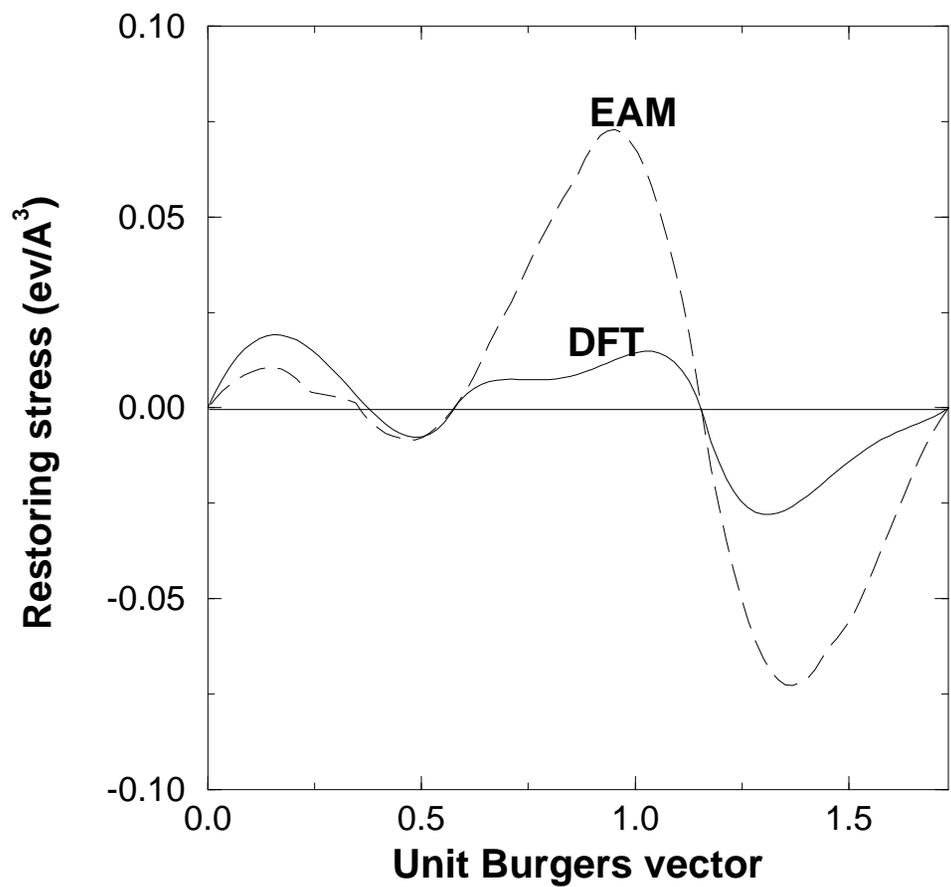
\begin{figure}
\caption{The restoring stresses along 
(a) [12$\bar{1}$] and (b) [101] (b) directions calculated 
from both the DFT and EAM GSF energy surfaces.}
\label{fig4}
\end{figure}

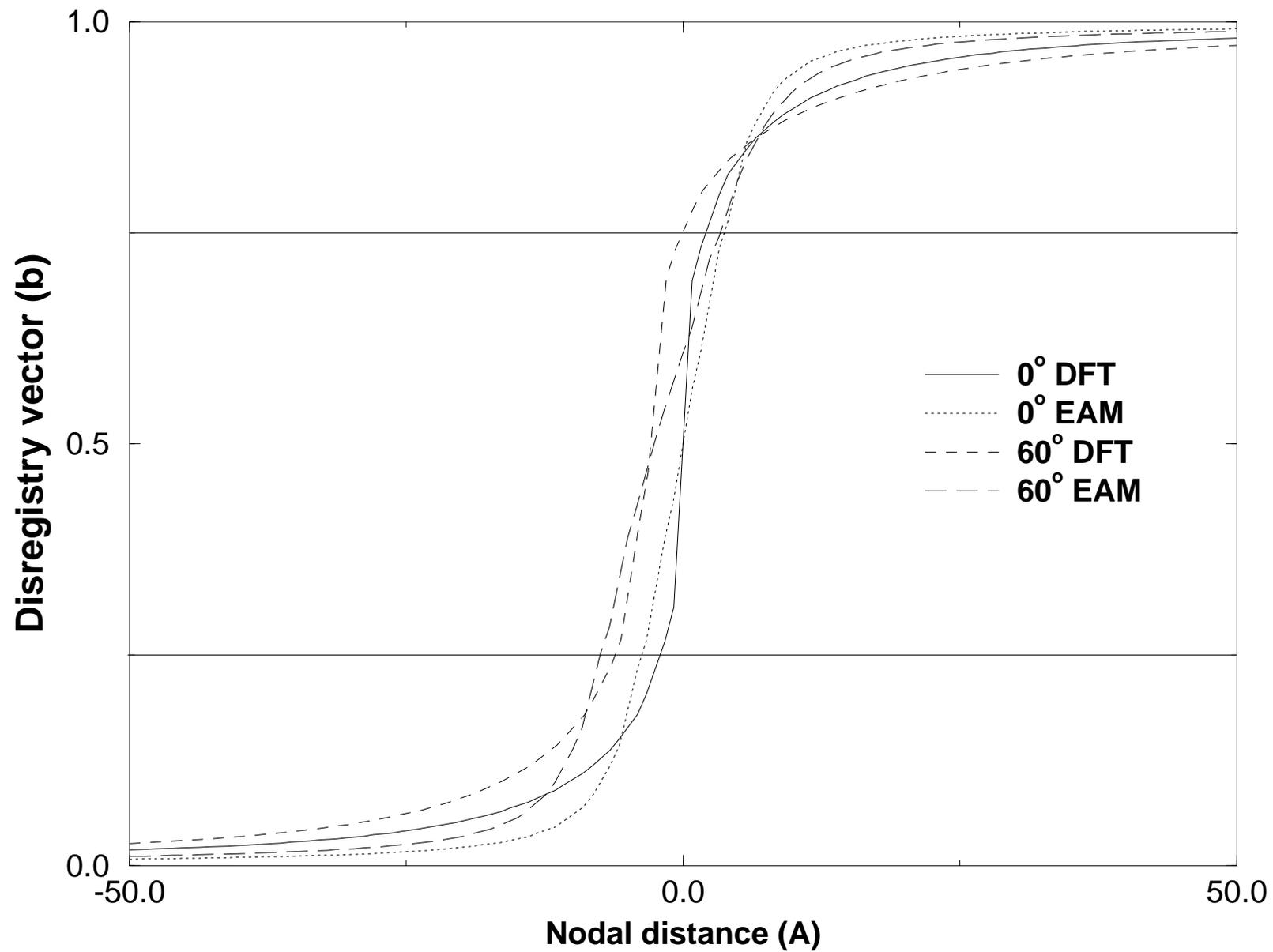
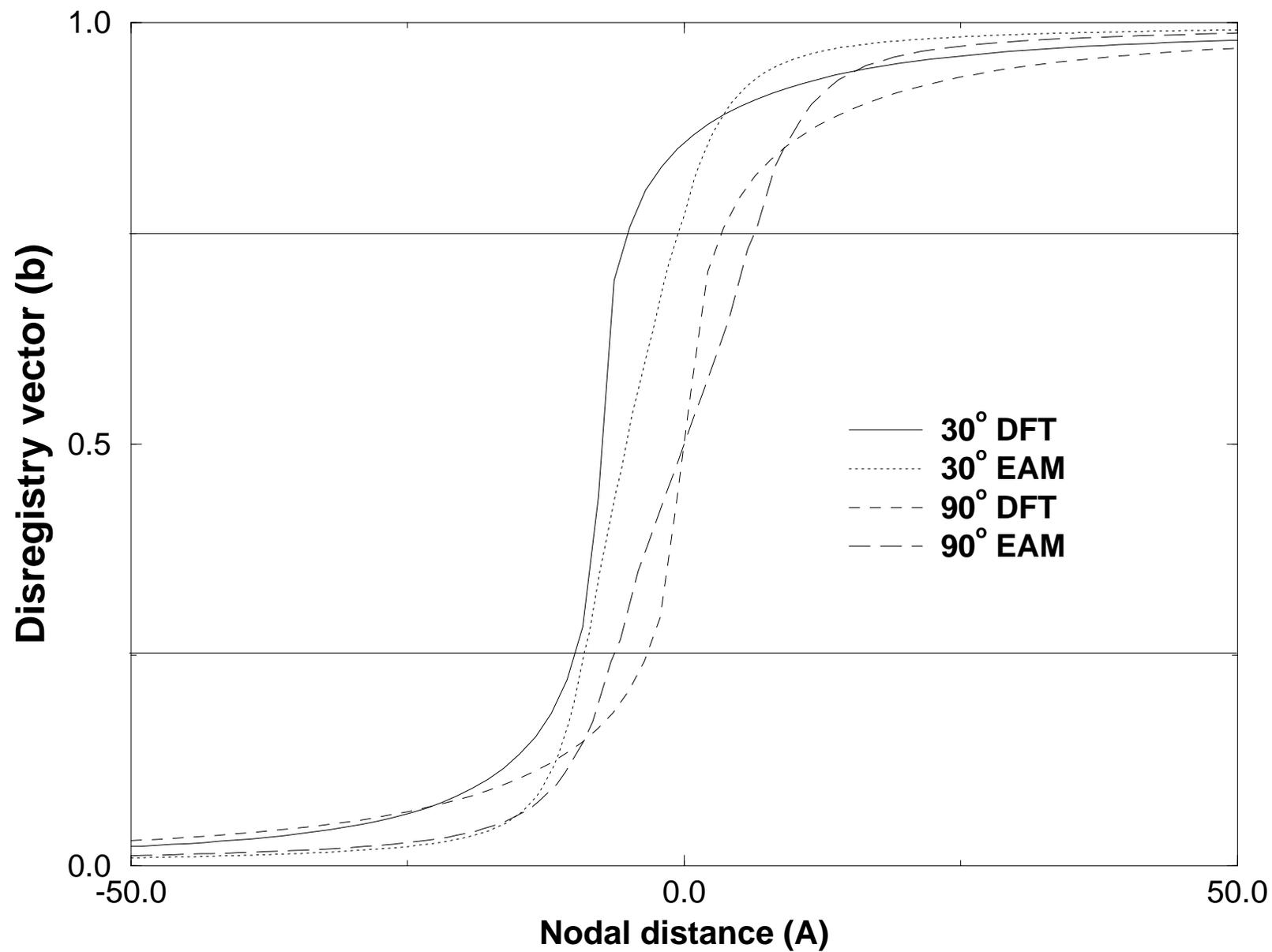
\begin{figure}
\caption{The disregistry vectors in unit of the Burgers vector 
obtained from the DFT and EAM calculations for: (a) the 
screw and 60$^\circ$ dislocations; 
(b) the 30$^\circ$ and edge dislocations.} 
\label{fig5}
\end{figure}

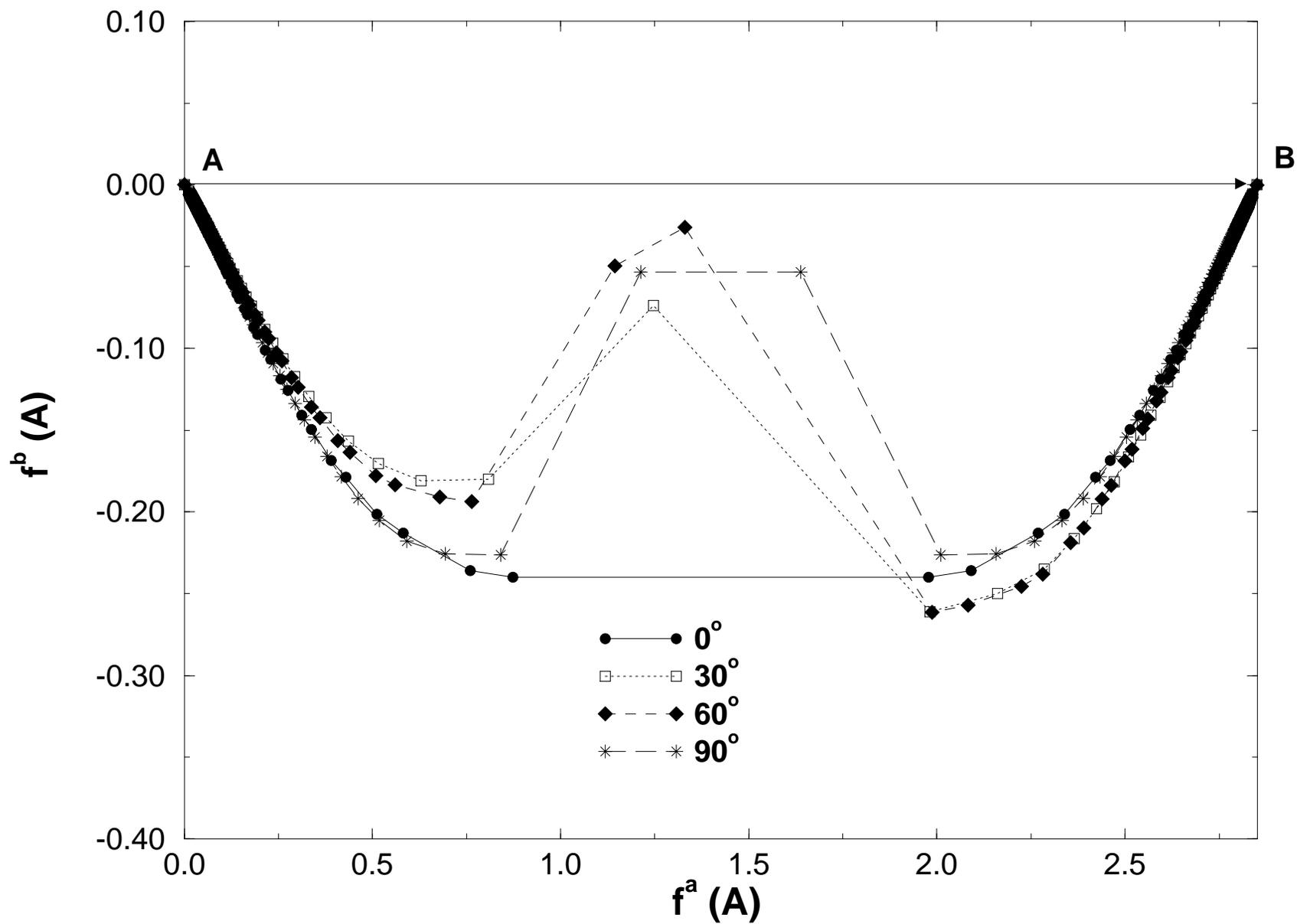
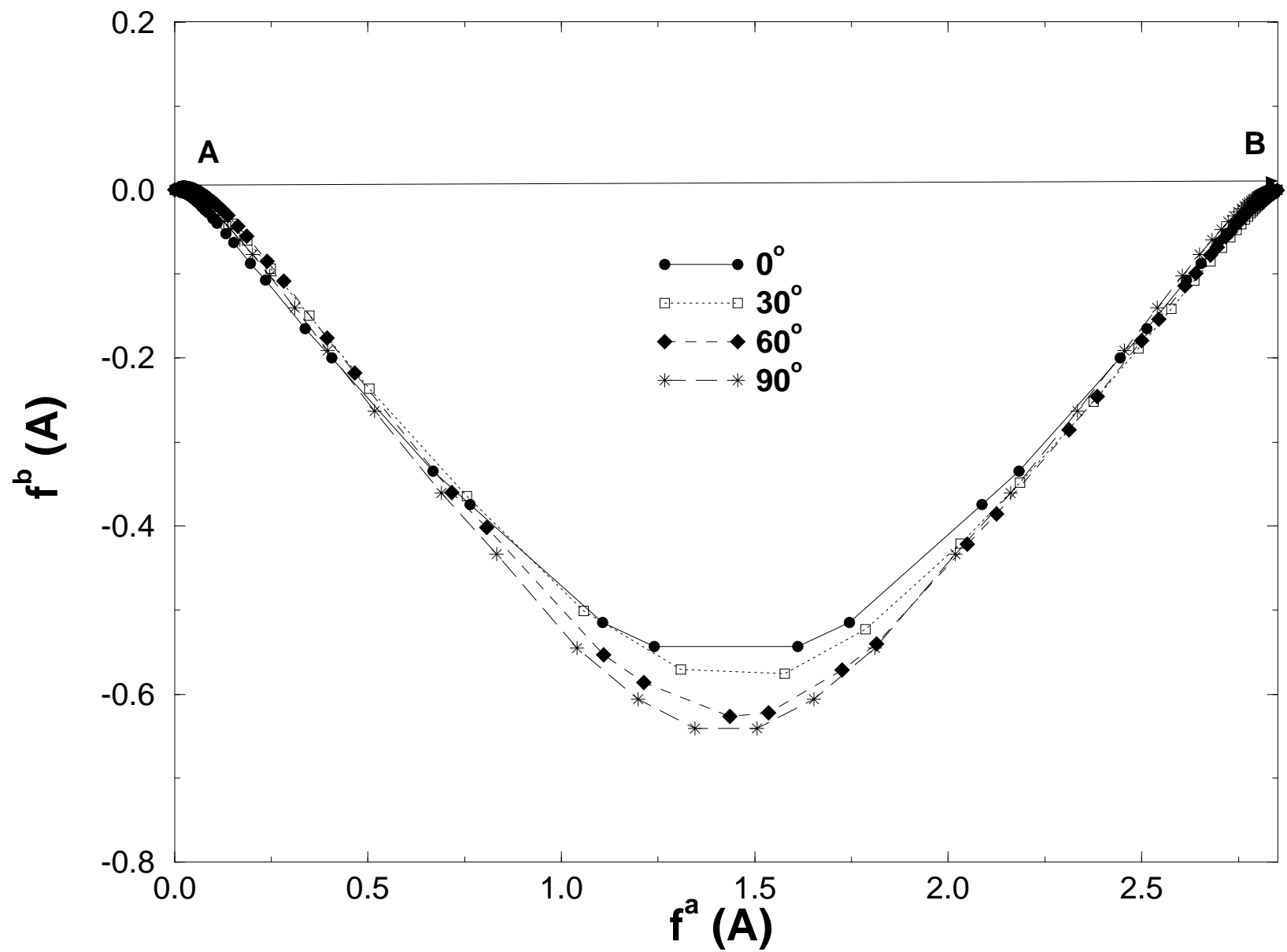
\begin{figure}
\caption{The disregistry paths for the four dislocations obtained from
(a) DFT calculations; (b) EAM calculations.} 
\label{fig6}
\end{figure}

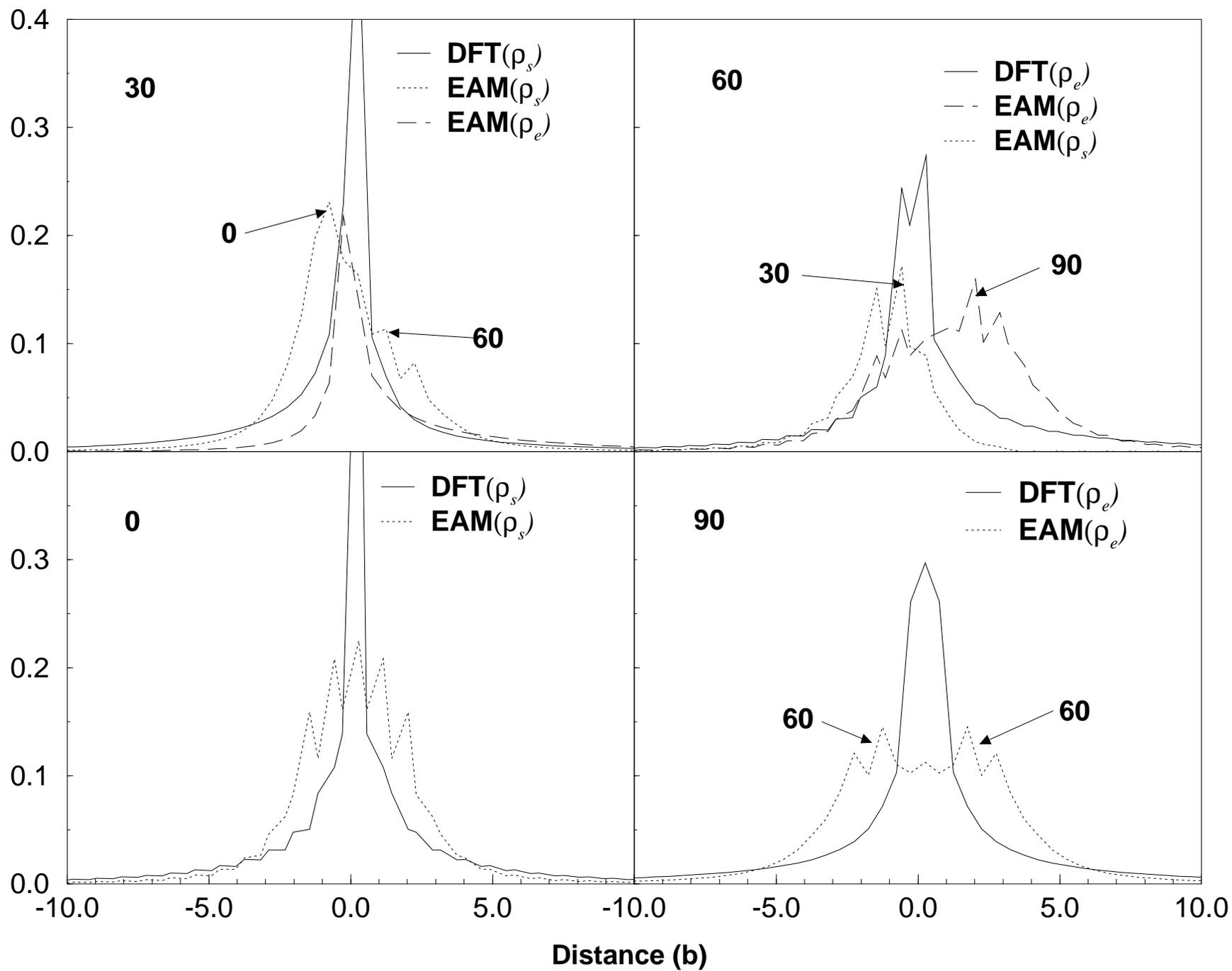
\begin{figure}
\caption{The density distributions for the four dislocations
(clockwise):  
screw, 30$^\circ$, 60$^\circ$ and edge, 
obtained by using DFT and EAM calculations:  
For the 30$^\circ$, 60$^\circ$ and edge dislocations, 
the characters of resultant partials 
within the EAM calculations are indicated (see also text).}
\label{fig7}
\end{figure}

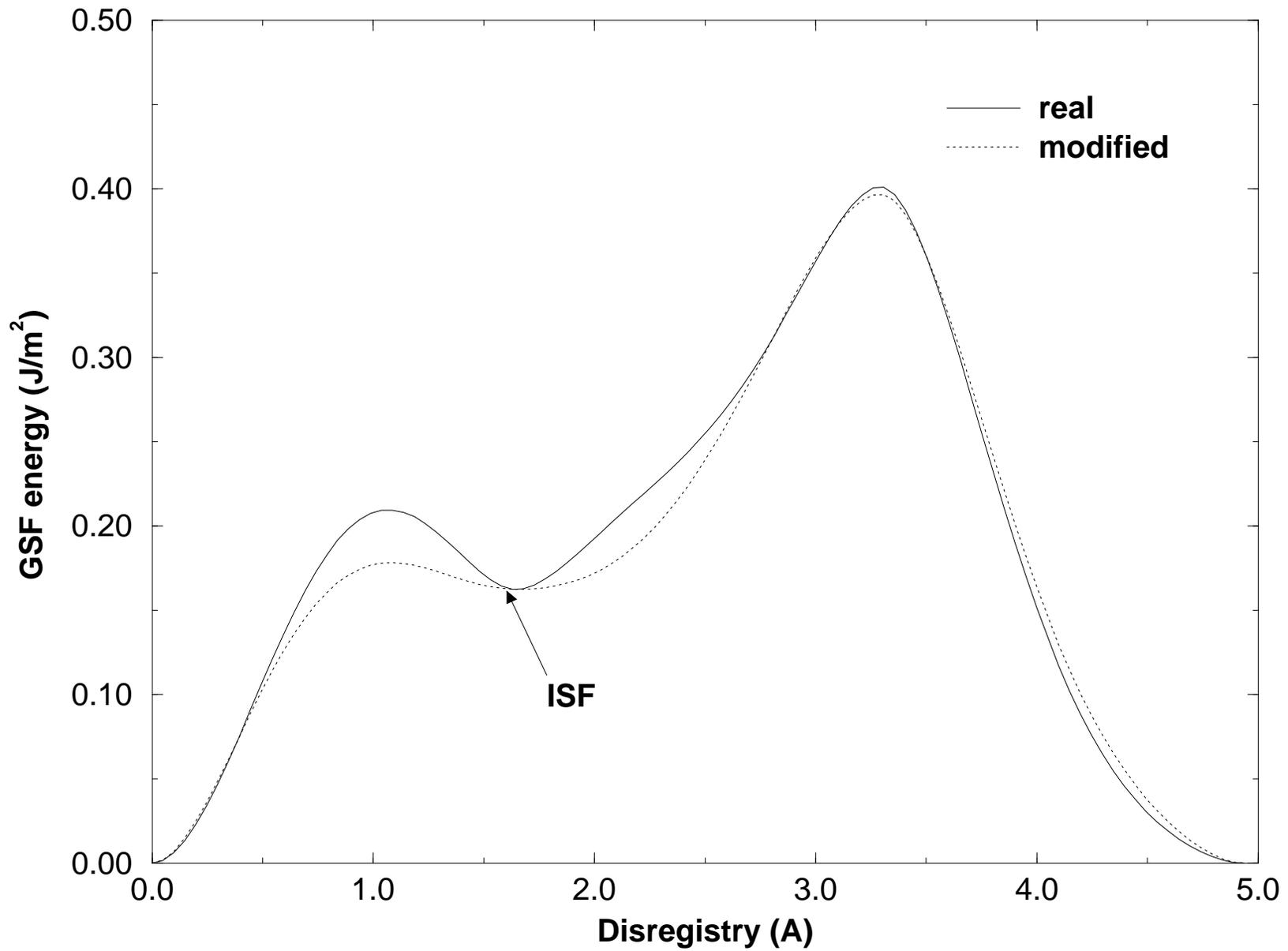
\begin{figure}
\caption{The projection of the DFT GSF energy surface along [12$\bar{1}$]
direction with and without modifying the real energy profile. The solid curve
represents the real energy profile without modification while the dashed 
curve represents the
modified GSF energy surface but keeping the same intrinsic stacking
fault energy as its real value.}
\label{fig8}
\end{figure}

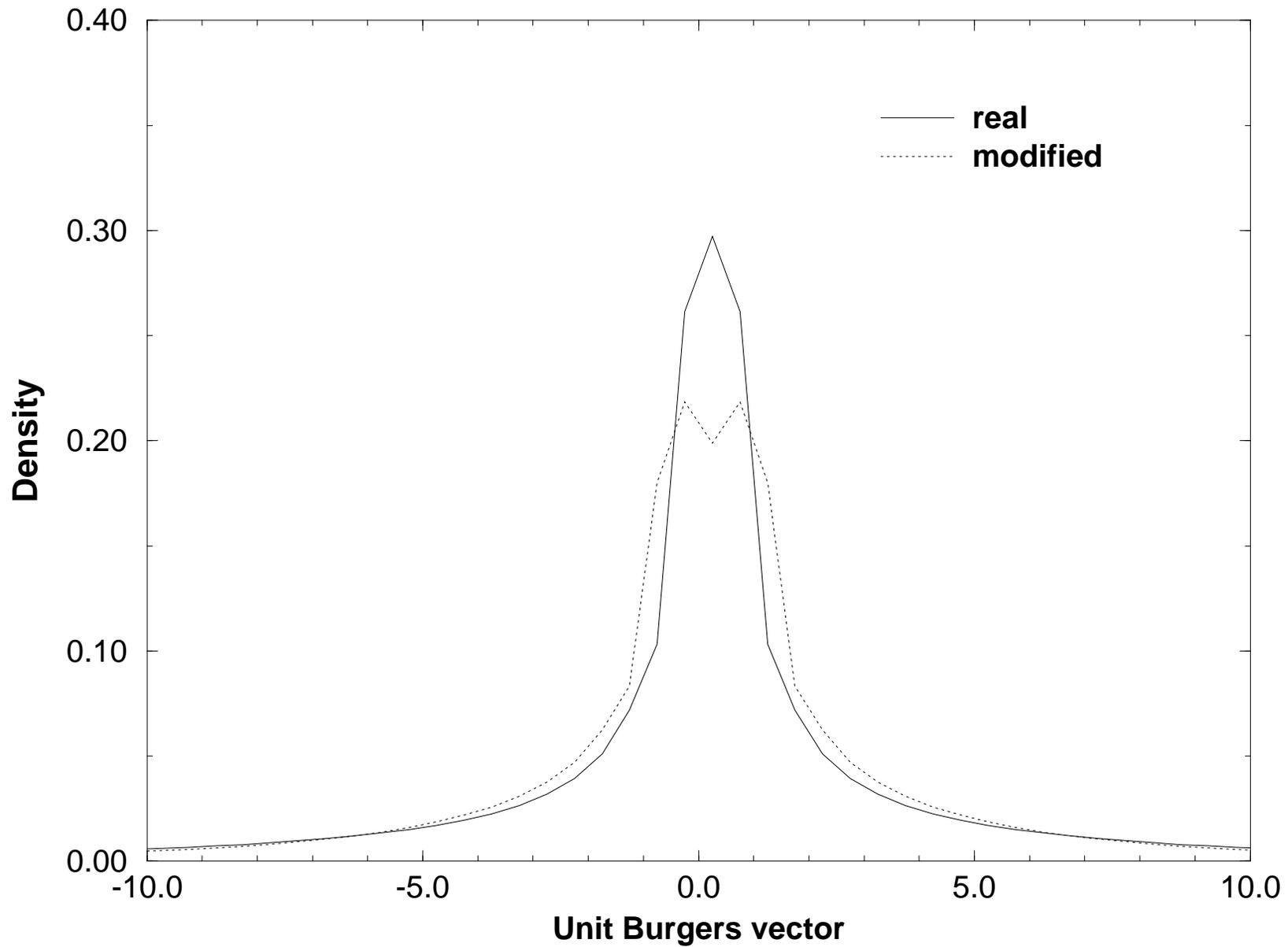
\begin{figure}
\caption{The dislocation density distribution for the edge dislocation
calculated
from the modified GSF energy surface (dashed line) and the real GSF energy
surface without modification (solid line). The distance is in terms of
the full Burgers vector.}
\label{fig9}
\end{figure}

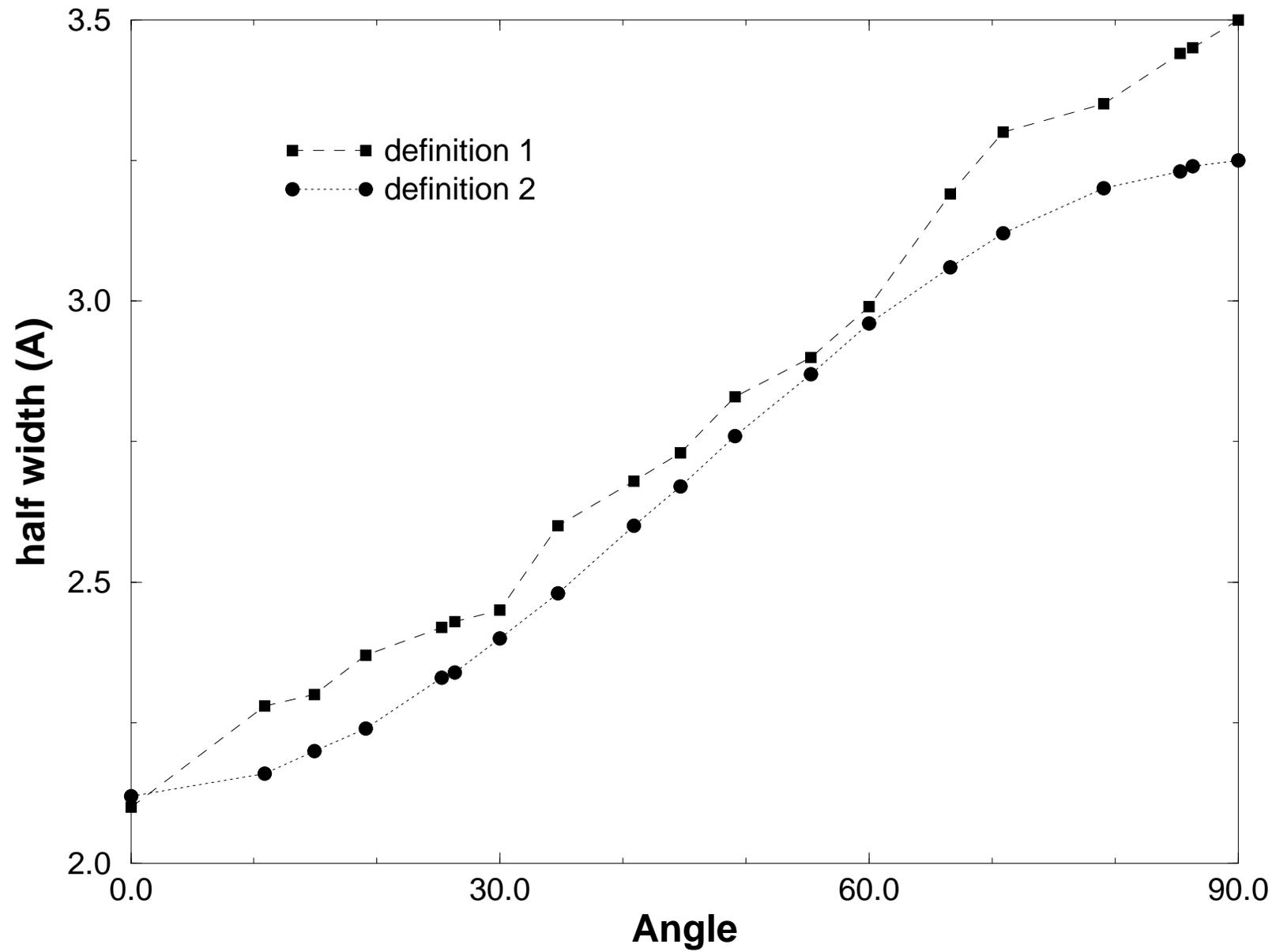
\begin{figure}
\caption{The half width (in \AA) 
as a function of dislocation 
orientation using the two different definitions. The square stands for the
definition through the disregistry vector and the dot stands for the definition
through the restoring stress (only DFT results shown).}  
\label{fig10}
\end{figure}

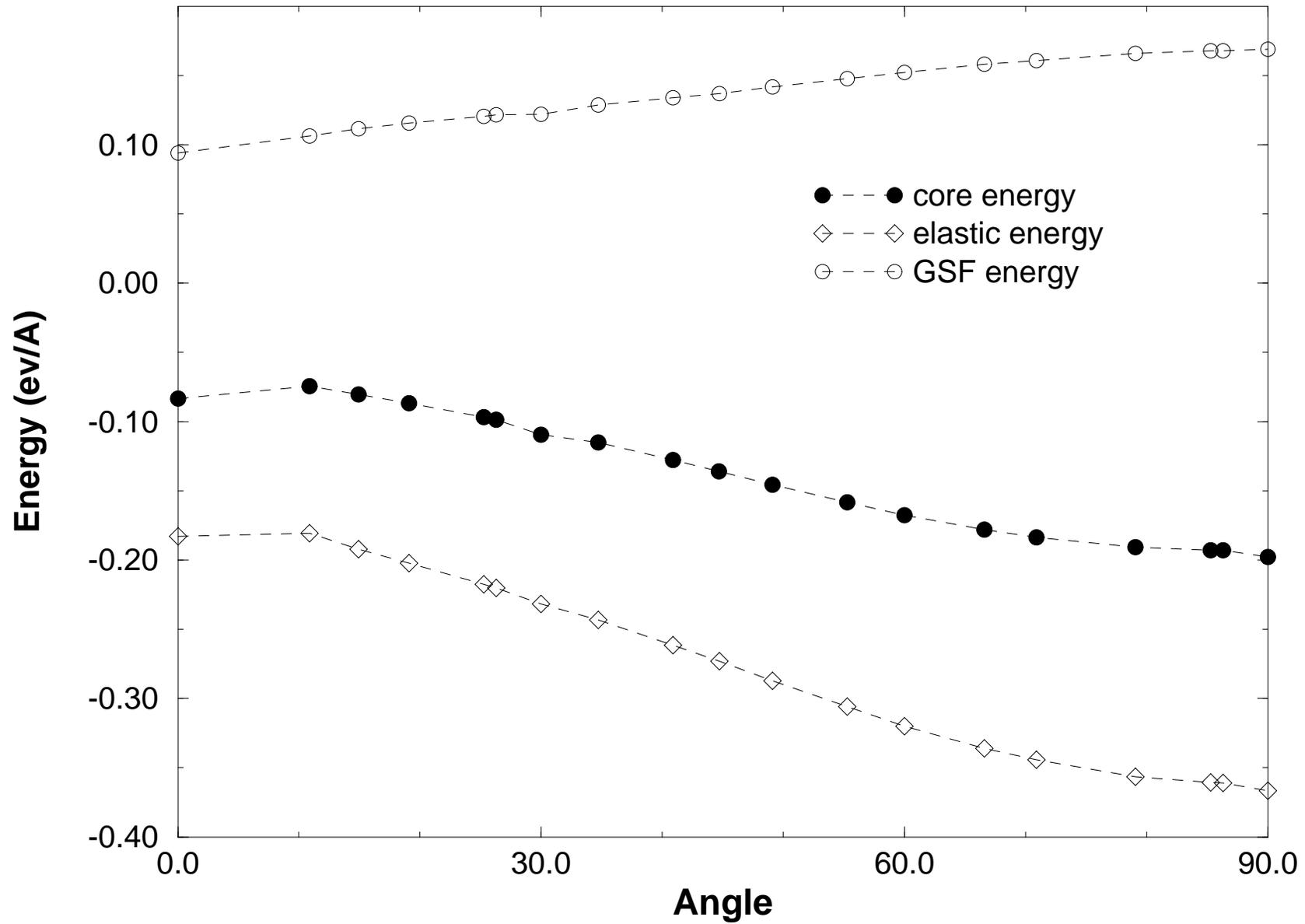
\begin{figure}
\caption{The core energy, 
elastic energy and GSF energy as a function of dislocation 
orientations.}
\label{fig11}
\end{figure}

\begin{figure}
\caption{The scaled Peierls stress as a function of 
the core width and the atomic spacing 
perpendicular to the dislocation line (the dashed line corresponds to
the formula from Jo\'{o}s {\it et al.}).}
\label{fig12}
\end{figure} 
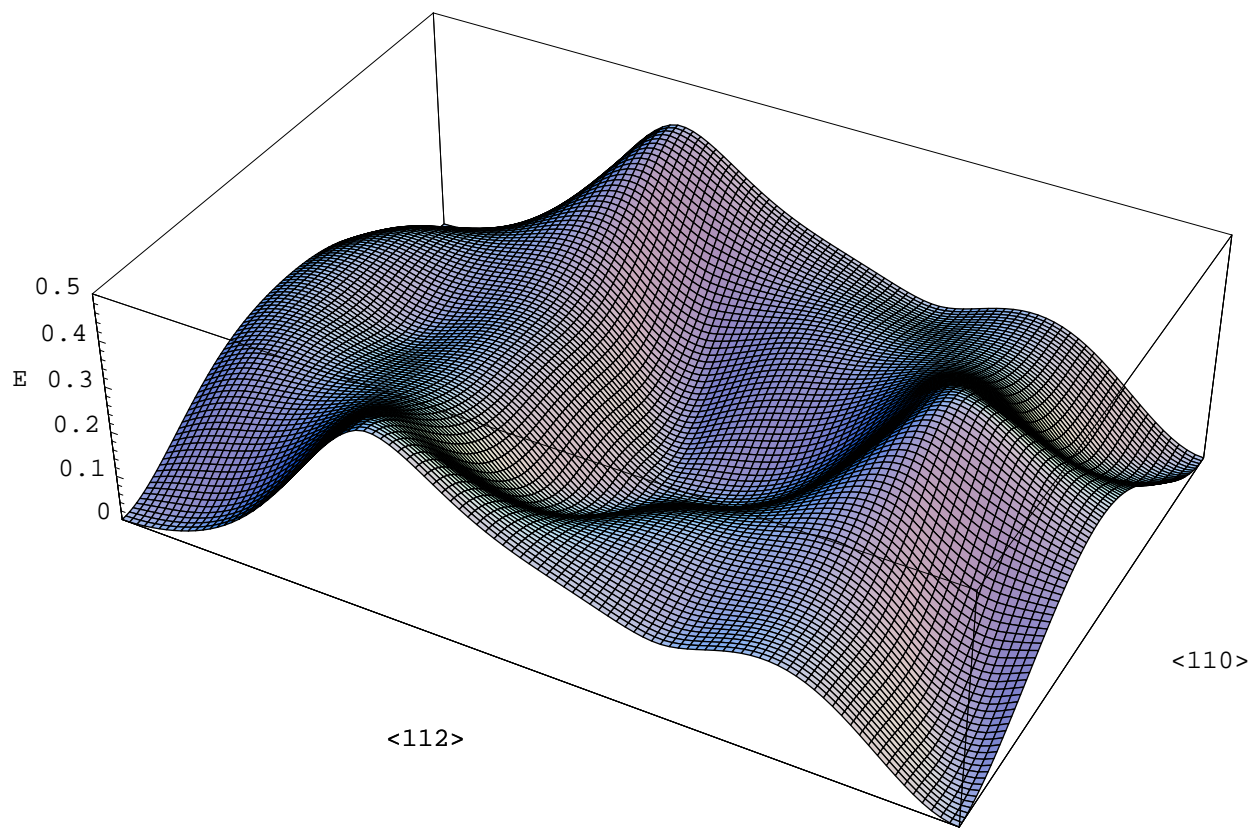
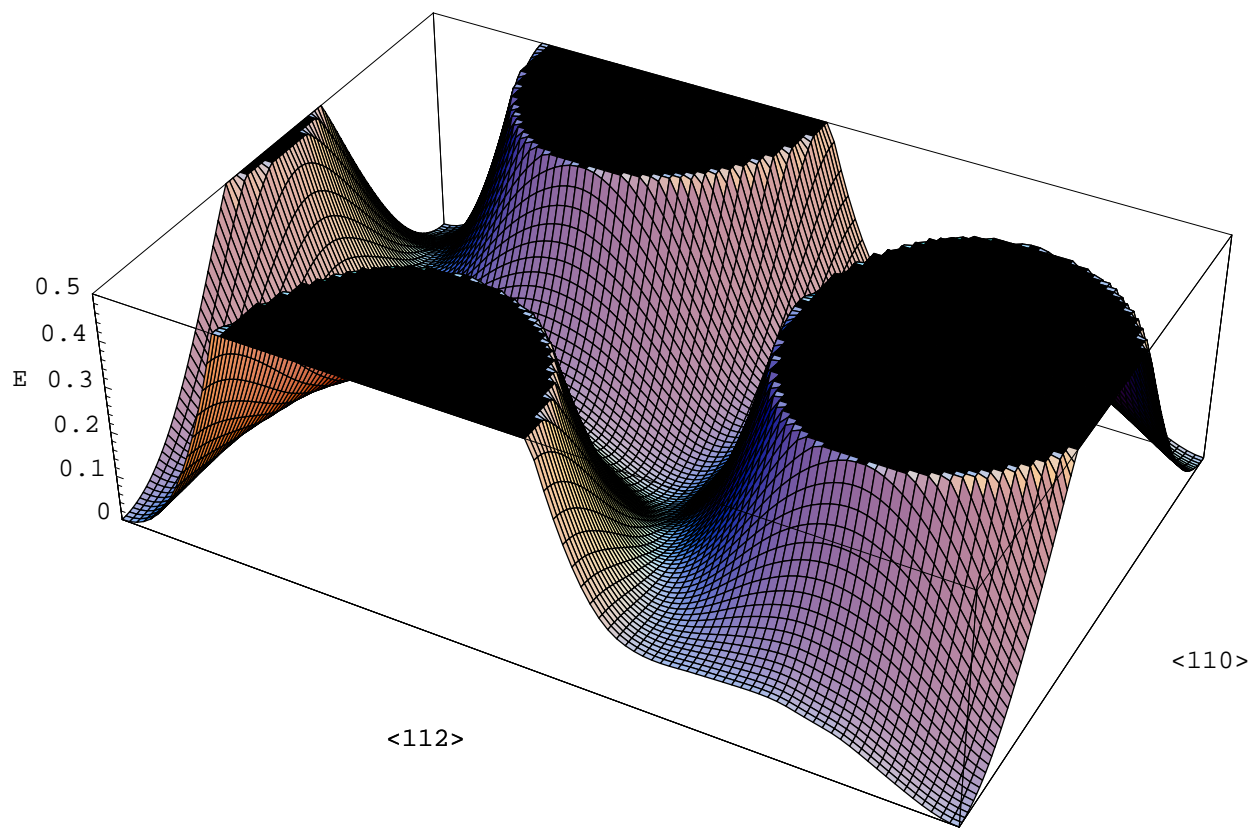
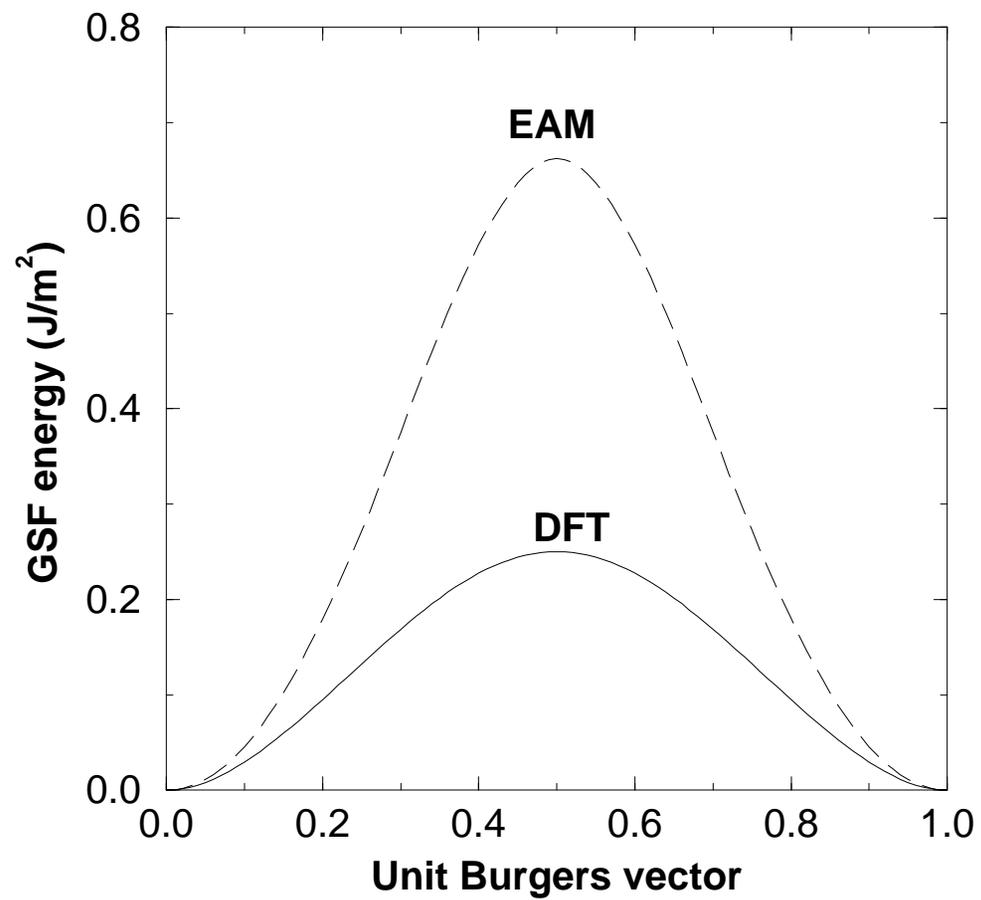
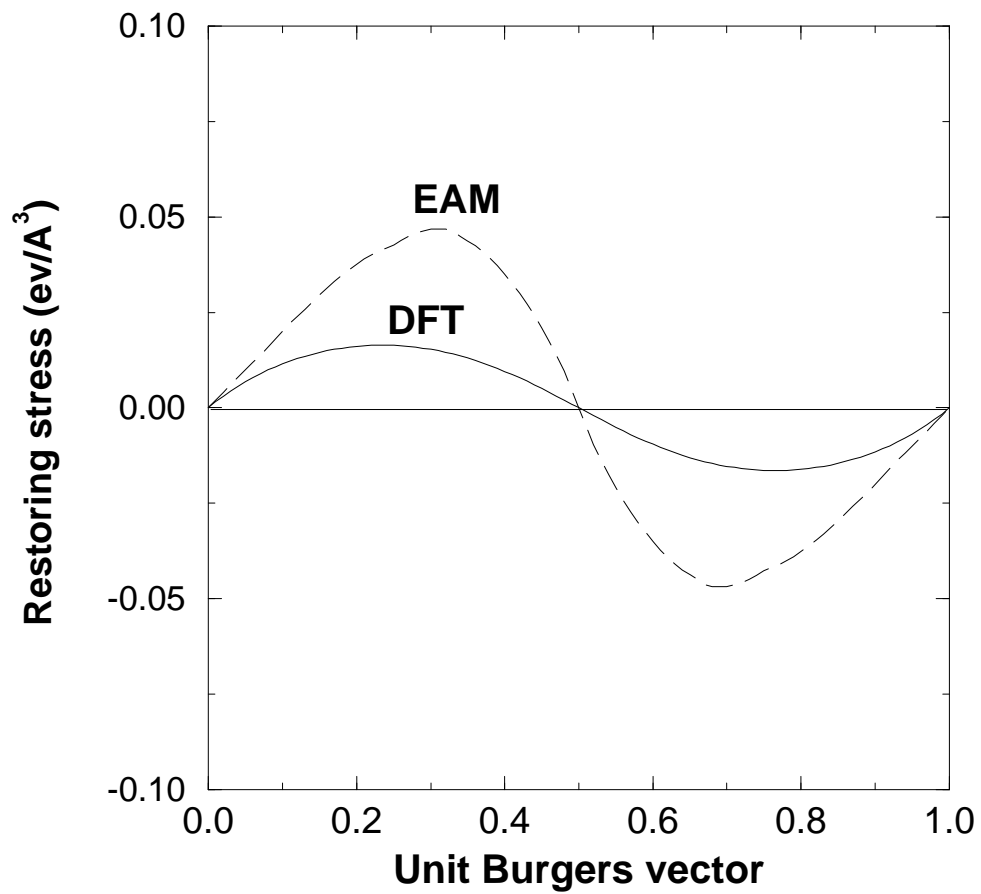
\end{document}